\pdfoutput=1

\documentclass[11pt]{article}
\usepackage[final]{acl}

\usepackage{times}
\usepackage{booktabs}
\usepackage{latexsym}
\usepackage[T1]{fontenc}
\usepackage[utf8]{inputenc}
\usepackage{microtype}
\usepackage{multirow}
\usepackage{algorithmic}
\usepackage[ruled,vlined]{algorithm2e}

\usepackage{url}
\usepackage{lscape}
\usepackage{amsmath}
\usepackage{dirtytalk}
\usepackage{color}
\usepackage{subfigure}
\usepackage{listings}
\usepackage{pgf-pie}
\usepackage[toc]{appendix}
\usepackage{xcolor}
\usepackage{framed}
\usepackage{amsfonts}
\usepackage{enumitem}
\usepackage{tikz}

\usepackage{soul}

\graphicspath{{./figures/}}

\newcommand\bsub[1]{\vspace{3pt}\noindent\textbf{#1}}

\usepackage{fancyhdr}
\fancypagestyle{firstpage}{
    \fancyhf{}
    \fancyhead[L]{Published as a conference paper at EMNLP 2024}
}
\begin{document}

\title{Sequential LLM Framework for Fashion Recommendation}

\author{
 \textbf{Han Liu\textsuperscript{$\spadesuit$}\textsuperscript{$\dagger$}},
 \textbf{Xianfeng Tang\textsuperscript{$\diamondsuit$}},
 \textbf{Tianlang Chen\textsuperscript{$\diamondsuit$}},
 \textbf{Jiapeng Liu\textsuperscript{$\diamondsuit$}},
\\
 \textbf{Indu Indu\textsuperscript{$\diamondsuit$}},
 \textbf{Henry Peng Zou\textsuperscript{$\clubsuit$}\textsuperscript{$\dagger$}},
 \textbf{Peng Dai\textsuperscript{$\diamondsuit$}},
 \textbf{Roberto Fernandez Galan\textsuperscript{$\diamondsuit$}},
\\
 \textbf{Michael D Porter\textsuperscript{$\diamondsuit$}},
 \textbf{Dongmei Jia\textsuperscript{$\diamondsuit$}},
 \textbf{Ning Zhang\textsuperscript{$\spadesuit$}},
 \textbf{Lian Xiong\textsuperscript{$\diamondsuit$}}
 \\
\textsuperscript{$\diamondsuit$}Amazon
 \textsuperscript{$\spadesuit$}Washington University in St. Louis
 \textsuperscript{$\clubsuit$}University of Illinois Chicago
 \\
 \small\texttt{\{h.liu1,zhang.ning\}@wustl.edu},
 \texttt{\{xianft,ctianlan,liujiape,indchand\}@amazon.com}, \\
 \small\texttt{pzou3@uic.edu}, 
 \texttt{\{pengdai,galanrob,mdporter,djia,lianxion\}@amazon.com}
 }

\maketitle

\begingroup\renewcommand\thefootnote{$\dagger$}
\footnotetext{Work done as an intern at Amazon.}
\endgroup

\begin{abstract}

The fashion industry is one of the leading domains in the global e-commerce sector, prompting major online retailers to employ recommendation systems for product suggestions and customer convenience. 
While recommendation systems have been widely studied, most are designed for general e-commerce problems and struggle with the unique challenges of the fashion domain.
To address these issues, we propose a sequential fashion recommendation framework that leverages a pre-trained large language model (LLM) enhanced with recommendation-specific prompts. Our framework employs parameter-efficient fine-tuning with extensive fashion data and introduces a novel mix-up-based retrieval technique for translating text into relevant product suggestions.
Extensive experiments show our proposed framework significantly enhances fashion recommendation performance.

\end{abstract}


\section{Introduction} \label{sec:intro}

In recent years, fashion e-commerce has garnered considerable global attentions from both consumers and investors.
By 2023, the U.S. retail fashion e-commerce market is projected to generate revenues exceeding 207 billion U.S. dollars \cite{fashion_sales}. One of the primary objectives of e-commerce is to provide a smooth purchase experience for consumers to purchase products they are looking for.
To this end, recommendation systems (RS) have become an essential part of many businesses \cite{zhang2019deep,jin2023amazon}. 
While existing fashion recommendation systems \cite{he2016vbpr,liu2017deepstyle,kang2017visually,yu2021visually} predominantly incorporate the visual appearance into the traditional recommendation, they often require resource-intensive processes for image collection and training. Additionally, they often struggle to capture the evolving nature of user interactions over time. In light of this, there has been a growing interest in sequential recommendation techniques \cite{sun2019bert4rec,kang2018self,li2023text}. These techniques model historical user interactions as temporally ordered sequences, thereby achieving remarkable efficacy in capturing both short-term and long-term user preferences.

While sequential recommendations have succeeded in general e-commerce, the fashion domain poses unique challenges. Our analysis of real-world user interactions on Amazon fashion highlights key differences:
First, the rapid fashion turnover leads to a sparse user-item interaction matrix, intensifying the cold-start problem \cite{liu2020heterogeneous}.
Second, extensive purchase comparisons demand sophisticated approaches to capture fine-grained user preferences.
Third, fashion-specific attributes like seasonality, occasion, and holiday trends require specialized modeling.
Fourth, diverse search queries that reflect explicit user intentions, necessitate novel modeling techniques. Beyond these fashion-specific challenges, traditional recommendation contexts often require specialized models tailored to particular scenarios, such as the cold-start problem \cite{dong2020mamo}, which will result in a large number of models that are challenging to maintain and scale.

To tackle these challenges holistically, we present a sequential fashion recommendation system augmented by a large language model (LLM). Trained on vast and diverse datasets, LLMs have a profound understanding of various domains. Leveraging their extensive knowledge and commonsense reasoning capabilities \cite{zhao2023survey}, LLMs provide a promising solution to generate meaningful recommendations. This is particularly beneficial in overcoming cold start problems and in accurately discerning fine-grained user preferences. Additionally, LLMs could offer a unified framework capable of addressing diverse recommendation tasks.
Our LLM-augmented recommendation system consists of three primary stages. In the first stage, prompt engineering techniques are used to devise specialized prompts that align with recommendation-specific goals, enabling LLM to perceive fine-grained user preferences.
In the second stage, we adapt Parameter-Efficient Fine-Tuning (PEFT) techniques \cite{hu2021lora, dettmers2023qlora} to mitigate prohibitively expensive training costs. 
In the final stage, we utilize predicted product titles and IDs to retrieve and rank potential candidate items. We present a mix-up-based retrieval technique that harnesses the strengths of both ID and title embeddings. 
Our contributions can be summarized as follows:

\vspace{-2.5mm}

\begin{itemize}[leftmargin=10pt,itemsep=0mm, parsep=0mm]

    \item We conduct an in-depth data analysis on real-world user interaction patterns, identifying four key characteristics for fashion recommendation.

    \item We propose a comprehensive recommendation framework tailored to the fashion domain. Within this framework, we propose advanced LLM enhancement techniques to address the unique challenges for fashion recommendation.

    \item The comprehensive evaluations demonstrate that the proposed framework significantly enhance recommendation performance.

\end{itemize}

\section{Related Work}  \label{related_work}

\bsub{Sequential Recommendation. }
Recommendation systems have gained significant interest from both academia and industry \cite{ma2022learning}, with sequential recommendation receiving particular attention due to its exceptional capabilities of capturing the long-term and short-term dynamics of users  \cite{li2022recguru,ma2023learning}. The objective of sequential recommendation is to predict the next items that users may be interested in based on their historical interactions. There are various techniques being proposed to model user sequential patterns, from the Markov Chain \cite{he2016fusing,rendle2010factorizing} in early works to recent neural network-based techniques, such as Gated Recurrent Units (GRU) \cite{hidasi2015session}, Convolutional Neural Network (CNN) \cite{tang2018personalized}, and Transformer \cite{sun2019bert4rec,kang2018self,hou2022core}. Recently, Recformer \cite{li2023text}, a transformer-based framework for learning transferable language representations, has been proposed for sequential recommendations. It has shown superior performance, especially in cold-start settings.

\bsub{Fashion Recommendation.}
Fashion recommendation systems, which target one vertical market - fashion and garment products, have gained popularity recently \cite{lin2019improving,hou2019explainable}. 
Existing approaches mainly use visual signals to capture fashion characteristics by enhancing item representations \cite{he2016vbpr,he2016fashionista,kang2017visually}, modeling visual compatibility \cite{chen2019pog,yin2019enhancing}, and identifying aesthetic and style information \cite{yu2021visually}.
For example, \citet{he2016vbpr} proposed the Visual Bayesian Personalized Ranking (VBPR), which incorporates visual features extracted from product images into matrix factorization frameworks using pre-trained CNNs. \citet{yin2019enhancing} utilized visual encodings to learn visual compatibility by training a triplet network, where an anchor item is paired with both a compatible and a non-compatible item to learn embeddings that capture visual compatibility. Additionally, \citet{yu2021visually} introduced a deep aesthetic network that extracts aesthetic features from product images, incorporating them into recommendations to model users' preferences for aesthetic appeal. 
The methods for extracting visual signals have evolved over time. Early studies typically used pre-trained CNNs for visual encodings \cite{he2016fashionista, he2016vbpr}. However, recent works have shifted towards training visual encoders on specialized datasets \cite{yin2019enhancing} or jointly training visual feature extractors and recommendation modules \cite{kang2017visually, lin2019improving}.

While incorporating visual signals is an inspiring direction, it falls outside the scope of and is furthermore orthogonal to our current study, which focuses on leveraging textual data to model user interactions. This choice is driven by the fact that learning effective product representations from images typically requires large datasets to generalize well \cite{deldjoo2022review}, which would introduce significant demands in terms of data collection and computational resources, making it challenging for industrial deployment.

\begin{figure}[t]
\centering
\includegraphics[width=0.875\linewidth]{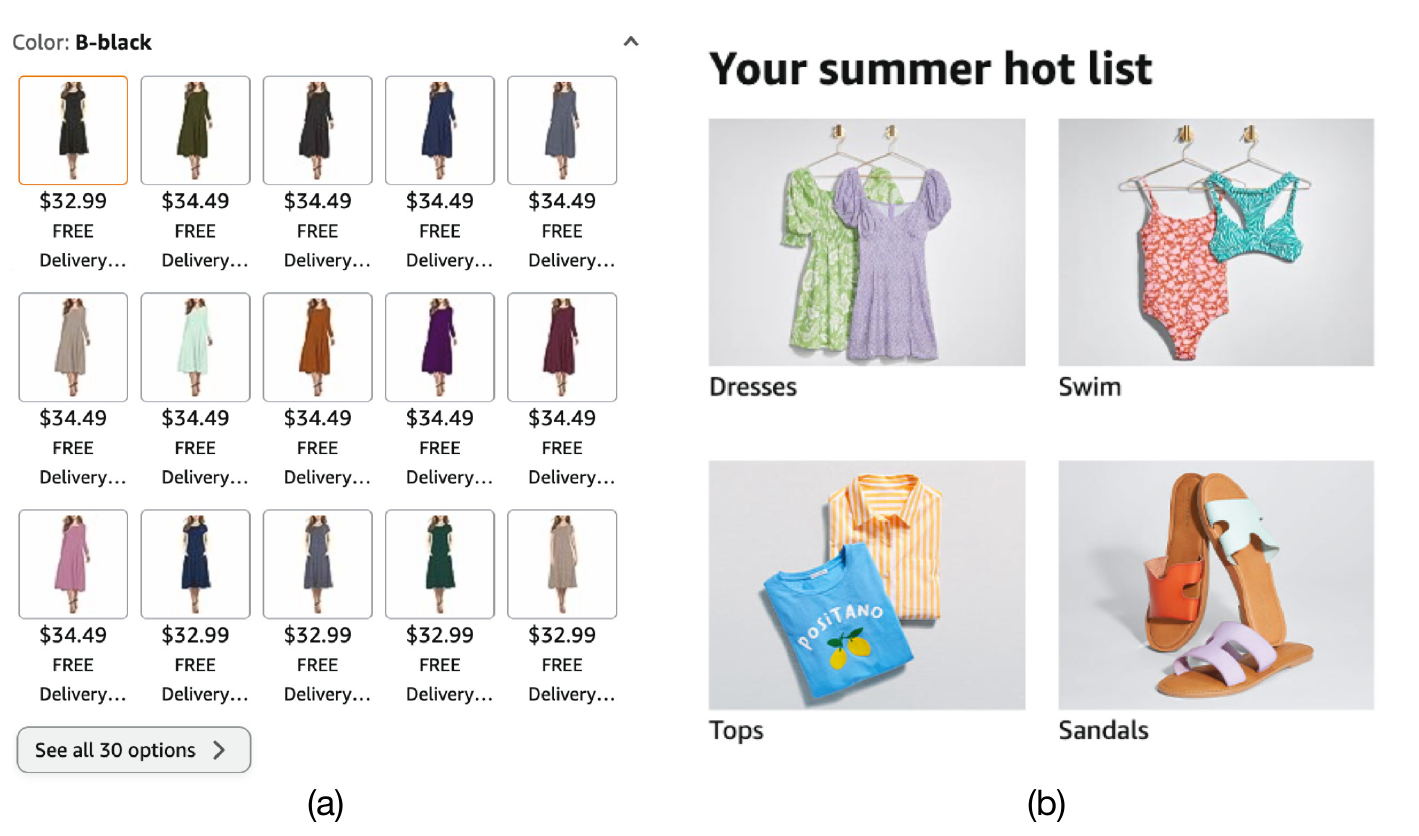}
\vspace{-1em}
\caption{Examples highlighting fashion characteristics. Figure (a) illustrates the extensive color variations in fashion products, while Figure (b) demonstrates the seasonality attributes of fashion items.}
\label{fig:fashion}
\vspace{-1em}
\end{figure}

\section{Fashion Characteristics} \label{sec:fashon_char}

Fashion-related shopping presents unique characteristics that must be carefully considered when developing RS. We conduct an in-depth analysis on real-user interaction patterns in Amazon Fashion,  
and identify the following key characteristics:

\vspace{2pt}\noindent\textit{C1: High Turnover of Products.} 
The fashion domain is characterized by a rapid turnover of items, introducing a continuous stream of unique new products to platforms. For instance, Amazon Fashion adds approximately 3 million new purchasable products each month. Additionally, the volume of new fashion items significantly exceeds that of other categories, being 3.6 times greater than new electronics and 6.7 times more than new toy products on Amazon. This constant influx leads to a notably sparse user-item interaction matrix, gives rise to the cold-start problem \cite{liu2020heterogeneous}.

\vspace{2pt}\noindent\textit{C2: Thorough Purchase Comparisons.}
    Users involved in fashion-related purchases tend to engage in more comprehensive comparisons than those shopping in other categories. For example,
    the average interaction length for fashion-related purchases is 55\% longer than for electronics and 81\% longer than for toy-related purchases. These comprehensive comparisons can be attributed to the extensive range of options—colors, styles, and sizes of fashion products. Figure \ref{fig:fashion} (a) provides an example of a typical shopping page for women's dresses, which offers 30 different color options.

\vspace{2pt}\noindent\textit{C3: Fashion Attribute-Driven Shopping.}
Fashion items often come with distinct attributes such as seasonality, occasion, and holiday-specific trends, which significantly influence user shopping intention. For instance, Figure \ref{fig:fashion} (b) shows a selection of items popular in summer, which might not receive the same attention in winter.

\vspace{2pt}\noindent\textit{C4: High Diversity of Search Queries.}
Search queries serve as a crucial context for understanding the evolving interests of users. We analyze the average volume of unique search queries over multiple days across three months. Our analysis shows that the number of unique search queries for fashion items is, on average, 2.63 times as much as electronics and 2.38 times as much as toys.

We highlight that while other industries may share some characteristics we've identified, the simultaneous presence of all four is unique to the fashion industry. Additionally, the importance of each characteristic in fashion differs from other domains. For example, attributes like seasonality, occasion, and trends have a more fine-grained influence on user choice in fashion compared to electronics or consumable products. In fashion, these factors influence not only availability but also social desirability and attractiveness at a given time.

\begin{figure}[t]
    \centering
    \includegraphics[width=0.47\textwidth]{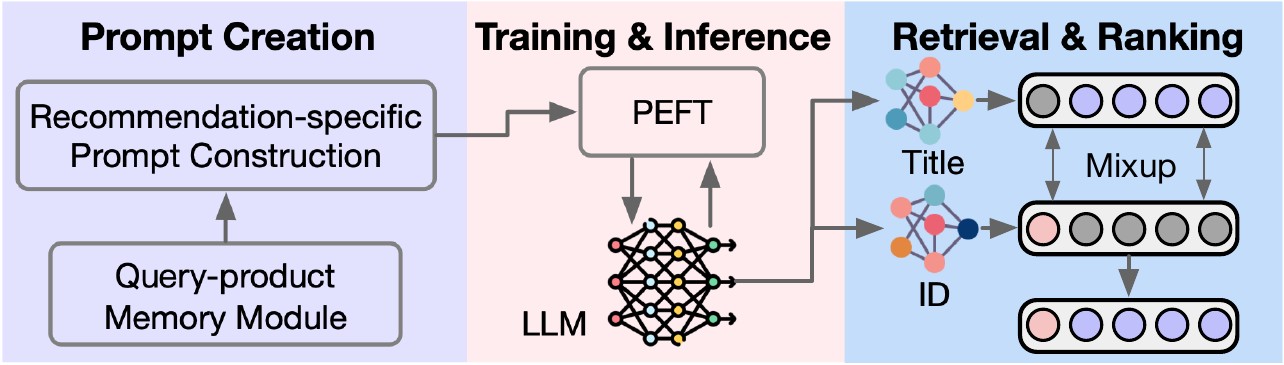}
    \caption{An overview of our method.}
    \vspace{-1em}
    \label{fig:overview}
\end{figure} 

\section{Method}  \label{method}

\subsection{Problem Formulation and Overview}   \label{subsec:overview}
\bsub{Problem Formulation.}
In the realm of sequential recommendation, consider a system composed of a set of users and items.  The set of users is represented by $\mathcal{U} = \{u_1,u_2,...,u_N\}$, the set of items by $\mathcal{V} = \{v_1,v_2,...,v_M\}$ and the set of queries as $\mathcal{Q} = \{q_1,q_2,...,q_S\}$. Each user $u_i \in \mathcal{U}$ is associated with an interaction sequence $S_i$, 
which can be denoted as $S_i=[(s_{1,i},a_{1,i},t_{1,i}),...,(s_{K,i},a_{K,i},t_{K,i})]$, where $K$ is the sequence length, $a_{k,i}$ represents the specific action type, $t_{k,i}$ represents the timestamp of the action, $s_{k,i}$ can be an item or a query depending on the action type. When $a_{k,i}$ represents a search action, $s_{k,i}\in \mathcal{Q}$ represents the $k$-th query. For other actions, $s_{k,i} \in \mathcal{V} $ represents the $k$-th interacted item. In this paper, we are interested in three action types, search, click, and purchase behavior, and we aim to predict the future item the user will be interested in purchasing after observing interaction sequences $S_i$. 
Additionally, each item $v$ is associated with an attribute dictionary containing various textual information, such as titles, colors, and descriptions. 
We formulate these as key-value attribute pairs and assign a unique ID to each item, in line with ID-based recommendation methods \cite{sun2019bert4rec, kang2018self}.

\bsub{Challenges and Overview.} 
Addressing the unique characteristics of fashion poses significant challenges for recommendation systems. For instance, the cold-start problem remains a persistent issue in recommender systems. Traditional approaches to mitigate this challenge often rely on complex and specialized architectures \cite{zhu2021learning,dong2020mamo}. Capturing fine-grained user preferences further complicates this task, typically requiring specialized modules \cite{wang2019dmfp, chen2021fg}. Additionally, leveraging search data to enhance recommendations remains relatively unexplored \cite{si2023search}, where the primary difficulty lies in the distinct nature of user intent in search versus recommendation tasks.
To address these challenges, we propose an LLM-augmented fashion sequential recommendation system, as shown in Figure \ref{fig:overview}. The process initiates with the creation of prompts. A query-product memory module assesses user-item interactions to identify top products associated with user queries. This information is synthesized into a natural language format using a recommendation-specific prompt template, incorporating fashion-related attributes. In the subsequent training stage, we utilize the prepared prompts to fine-tune the LLM through a Parameter-Efficient Fine-Tuning method. 
Finally, in the retrieval and ranking stage, we convert the generated titles and IDs into embeddings using two specialized models. These embeddings are integrated into a retrieval module with a mixup strategy, obtaining the final recommended items. 

\begin{figure}[t]
\centering
\includegraphics[width=0.95\linewidth]{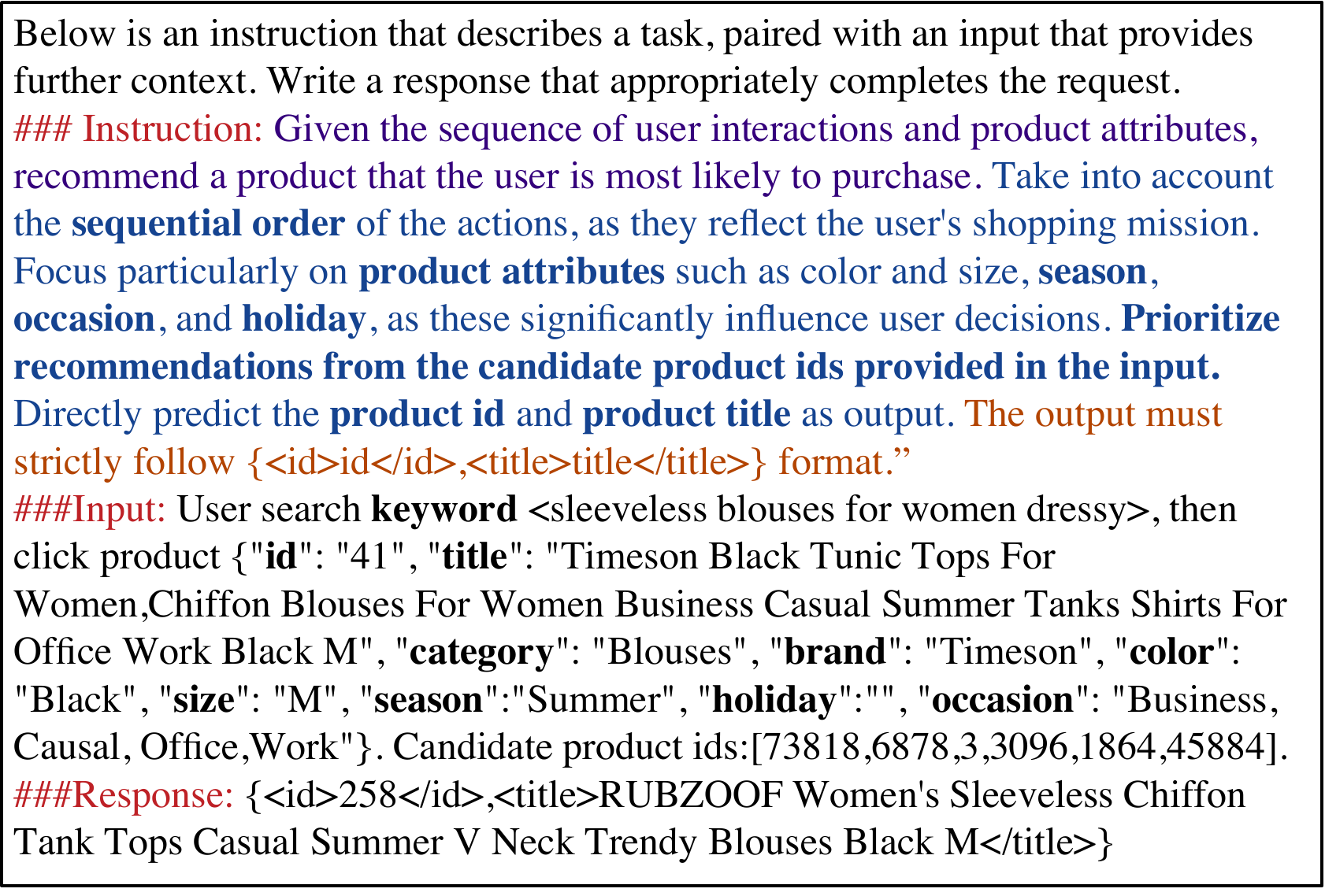}
\caption{The demonstration example of our prompt.}
\vspace{-1em}
\label{fig:prompt_exp}
\end{figure}

\subsection{Prompt Design}  \label{subsec:prompt_design}

\bsub{Recommendation-specific Prompt Construction. }
Prompting offers a natural and intuitive interface for humans to interact with LLMs \cite{zhou2022large}. 
Given that LLMs are initially trained for general tasks, specialized prompts are essential for aligning LLMs with recommendation-specific goals. 
A demonstration example of our designed prompt is given in Figure \ref{fig:prompt_exp}.

The instruction segment aims to clearly define the task and consists of three core elements: \textit{task description} (highlighted in purple), \textit{execution requirement} (highlighted in blue), and \textit{format indicator} (highlighted in brown). In the task description, we explicitly specify that the context is a recommendation task. The execution requirement emphasizes a set of strategies tailored to address the unique characteristics of fashion. A prime emphasis here is the consideration of sequential order. To address \textit{C2}, we intend for the LLM to focus on varying attributes, as they offer insight into users' fine-grained preferences. To address \textit{C3}, we emphasize the importance of fashion-specific attributes. To address \textit{C4}, we observed that customers generally have preferences in purchasing the top exposure results on the shopping page, thus we instruct the LLM to prioritize the recommendation in the top exposure results corresponding to the search query. Finally, the format indicator strictly defines an output format for automated decoding. 
The input segment is a refined representation of user-item interactions, enriched with detailed item attributes. Search queries are also included to highlight their importance in the recommendation task. 
The response segment, employed only during the training phase, encapsulates the final item purchased by the user, including both the product ID and title.

\bsub{Query-product Memory Module. }  
We observe that users frequently opt to purchase items listed at the top of their search results. In response to this behavior, we propose a Query-Product Memory Module that preserves key-value pairs consisting of search queries and their corresponding product listings. To obtain these product lists, data is grouped by search queries and then sorted by organic position. Recognizing that queries can appear in various forms that convey similar meanings, we employ CLIP \cite{radford2021learning} to convert these queries into embedding vectors, which serve as the keys in our module. During the recommendation process, the current search query is transformed into its respective embedding, enabling us to compute the cosine distance, identify the nearest $Q$ matching queries, and subsequently retrieve their associated top $V$ products.

\subsection{Training Strategy} \label{subsec:train_inference}

Low-Rank Adaptation (LoRA) \cite{hu2021lora} has emerged as a notable Parameter-Efficient Fine-Tuning (PEFT) technique, offering performance comparable to full fine-tuning while requiring substantially fewer trainable parameters. Consequently, we have adopted this method to fine-tune our model.
We further reduce memory usage by employing model quantization as implemented in QLoRA \cite{dettmers2023qlora}. Specifically, we maintain distinct storage and computation data types. We quantize the model to a more memory-efficient storage type and, during the forward and backward passes, dequantize the data back to the computation type to avoid performance loss.
During our preliminary experiments, we observed that the model exhibited a 7\% likelihood of generating outputs in an inconsistent format. This inconsistency made the automated decoding of product IDs and titles challenging.
One possible reason is that product titles in e-commerce often display limited sentence coherence and are more like a collection of individual words, setting them apart from typical natural language structures. To mitigate this issue, we identified the high-perplexity prompts and subjected them to additional training cycles relative to their lower-perplexity counterparts.

\subsection{Retrieval and Ranking Method} \label{subsec:retrieval_rank}

\bsub{Title Embedding Model. } 
To effectively capture the semantic similarity of item titles in recommendation tasks, we leveraged the insight that the items that were purchased in the same search queries should be similar in embedding space. Based on this insight, we first tokenize both the query and product title. 
Once tokenized, the model computes the embeddings for query and title by employing the LSTM model.
We train the whole model using a triplet loss \cite{jiang2016deep}, where we pair two hard negatives with one positive pair during each forward pass. The positive match means the item that was purchased from the query. We choose the title that is closest to the query but is not a positive match and the query that is closest to the title but is not a positive match as the hard negatives.

\bsub{ID Embedding Model. } 
The ID embedding model maps pre-defined item IDs into their embeddings. We leverage the item embedding table from the CORE model, which has demonstrated superior performance compared to state-of-the-art methods. Specifically, we train the CORE model using user-item interaction sequences, then keep only the item embedding table as our ID embedding model.

\bsub{Retrieval with Mixup. } 
After obtaining both title embeddings and ID embeddings, the next step is to perform retrieval and ranking processes to get the candidate items for recommendation. Title embeddings are designed to capture the semantic content of an item's title, thus offering better generalization, even if an item hasn't been seen before (\textit{i.e.} cold start). 
Conversely, ID embeddings are designed to uniquely represent specific items, so the embedding can capture nuances specific to that item, thus being suitable in top matches. 
To effectively combine the advantages of the two methods, we propose a mixup-based retrieval method. This approach begins with separate retrievals based on title and ID embeddings, resulting in two distinct lists of items. To generate our final list of top-$K$ items, we adopt the following approach: We select the top-$N$ items from the ID embedding-based list. Subsequently, we choose items ranging from positions $N+1$ to $K$ from the title embedding-based list. We set $N=1$ for all our experiments.

\section{Experiments} \label{sec:experiment}

\subsection{Experimental Setup} \label{subsec:exp_setup}

\bsub{Datasets. }
We have collected a large-scale dataset derived from customer interactions on the Amazon fashion service,  containing approximately 5.9 million user shopping interactions with a total of 2.4 million products.
We aggregated them into four primary categories: Luggage and Bags, Footwear, Accessories and Jewelry, and Clothing. The sequences included three action types: search, click, and purchases. We also filtered the 'click' interactions on the items that were eventually purchased.
Each item in our dataset is described by an array of attributes such as item and user identifier, product title, category, brand, color, and size. 
The statistics of the data after processing are given in Table \ref{tab:data_stat}.

\begin{table}[htbp]
\small
\caption{Statistics of the datasets. Avg. Len. represents the average length of interaction sequences.}
\vspace{-1em}
\label{tab:data_stat}
\renewcommand{\arraystretch}{1.2}
\setlength{\belowrulesep}{1.5pt}
\setlength{\tabcolsep}{0.1pt}
\begin{tabular}{c|c|c|c|c|c}
\toprule
\hline
\textbf{Dataset}         & \textbf{\#Users} & \textbf{\#Items} & \textbf{\#Inters.} & \textbf{Avg. Len.}   & \textbf{Density}  \\ \hline
Lug. \& Bags & 10,611   & 61,550   & 131,647    & 12.41 & 2.02E-04 \\ \hline
Footwear        & 63,273   & 380,385  & 714,628    & 11.29 & 2.97E-05 \\ \hline
Acc. \& Jew.  & 106,104  & 524,433  & 1,376,999   & 12.98 & 2.47E-05 \\ \hline
Clothing        & 274,285  & 1,386,910 & 3,635,414   & 13.25 & 9.56E-06 \\ \hline
\end{tabular}
\end{table}

\begin{table*}[t]
\scriptsize
\centering
\caption{Performance comparison of our method with the state-of-the-art methods across different datasets. }
\vspace{-1em}
\label{tab:all_results}
\renewcommand{\arraystretch}{1.5}
\setlength{\belowrulesep}{1.4pt}
\setlength{\tabcolsep}{2.4pt}
\begin{tabular}{c|ccc|ccc|ccc|ccc}
\toprule
\hline
\multirow{2}{*}{\textbf{Method}} & \multicolumn{3}{c|}{\textbf{Luggage \& Bags}} & \multicolumn{3}{c|}{\textbf{Footwear}} & \multicolumn{3}{c|}{\textbf{Accessories and Jewelry}} & \multicolumn{3}{c}{\textbf{Clothing}} \\ \cline{2-13} 
                        & Recall@10    & NDCG@10    & MRR      & Recall@10  & NDCG@10 & MRR    & Recall@10       & NDCG@10      & MRR         & Recall@10 & NDCG@10 & MRR    \\ \hline
GRU4Rec       & 0.0336       & 0.0221     & 0.0185   & 0.0185     & 0.0124  & 0.0105 & 0.0230          & 0.0155       & 0.0131      & 0.0221    & 0.0155  & 0.0134 \\ \hline
SASRec         & 0.1015       & 0.0613     & 0.0487   & 0.0857     & 0.0548  & 0.0452 & 0.1440          & 0.0825       & 0.0631      & 0.1485    & 0.0827  & 0.0617 \\ \hline
BERT4Rec      & 0.0975       & 0.0600     & 0.0485   & 0.1405     & 0.0770  & 0.0568 & 0.0904          & 0.0580       & 0.0479      & 0.0676    & 0.0435  & 0.0361 \\ \hline
NextItNet     & 0.0176       & 0.0094     & 0.0069   & 0.0123     & 0.0103  & 0.0097 & 0.0185          & 0.0150       & 0.0139      & 0.0111    & 0.0087  & 0.0079 \\ \hline
CORE         & 0.2612       & 0.1404     & 0.1027   & 0.3075     & 0.1566  & 0.1092 & 0.2493          & 0.1327       & 0.0962      & 0.1989    & 0.1008  & 0.0699 \\ \hline
Recformer      & 0.2577       & 0.1692     & 0.1455   & 0.2181     & 0.1352  & 0.1161 & 0.1719          & 0.1132       & 0.1004      & 0.1741    & 0.1102  & 0.0972 \\ \hline
Recformer w/ query      & 0.2642       & 0.1834     & 0.1524   & 0.2325     & 0.1436  & 0.1225 & 0.1990          & 0.1220       & 0.1046      & 0.1888    & 0.1276  & 0.1059 \\ \hline
Ours                    & \textbf{0.2786}       & \textbf{0.2037}     & \textbf{0.1804}   & \textbf{0.3377}     & \textbf{0.1791}  & \textbf{0.1470} & \textbf{0.2624}          & \textbf{0.1676}       & \textbf{0.1343}      & \textbf{0.2593}    & \textbf{0.1658}  & \textbf{0.1440} \\ \hline
\end{tabular}
\vspace{-1.5em}
\end{table*}

\bsub{Evaluation Settings. }
To assess the efficacy of our sequential recommendation approach, we employ three widely used metrics: Recall@N, NDCG@N, and MRR, where N is set to 10.  During the evaluation, we rank the ground-truth item (\textit{i.e.}, final purchased item) of each sequence among all items in the same category and report the average values of all sequences in the test data. We employ the common leave-one-out strategy \cite{sun2019bert4rec,zhou2020s3} to split the data for evaluation.

\bsub{Baselines. }
To evaluate the performance of the proposed method, we compare it with the following representative baselines: GRU4Rec \cite{hidasi2015session}, SASRec \cite{kang2018self}, BERT4Rec \cite{sun2019bert4rec}, NextItNet \cite{yuan2019simple}, CORE \cite{hou2022core}, and Recformer \cite{li2023text}. To ensure a more fair comparison, we also compare with Recformer w/ query, which is similar to Recformer, with the only change of adding the search query as part of the input. 

\bsub{Implementation Details. }
We select Falcon-7b \cite{falcon} as our base LLM model. We implemented the training framework by using Huggingface PEFT library\footnote{PEFT: \href{https://huggingface.co/docs/peft/index}{https://huggingface.co/docs/peft/index}}. For LoRA, we set rank $r$ to 16, scaling parameter $\alpha$ to 16, and dropout rate to 0.05. The maximum number of tokens for each interaction sequence is 1024. The models were trained on 8 Nvidia Tesla V100 GPUs. We optimized Falcon with AdamW optimizer \cite{loshchilov2017decoupled} with a learning rate 2e-5. We only fine-tuned the model for 1 epoch, except in cases involving prompts with high perplexity, where we selected the top 20\% of prompts as high perplexity prompts for fine-tuning 3 epochs. During the generation, we set the max new tokens to 64, the temperature to 0.05, and the probability threshold of nucleus sampling to 0.95. For the implementation of Recformer, we followed the official Github repository\footnote{Recformer: \href{https://github.com/JiachengLi1995/Recformer}{https://github.com/JiachengLi1995/Recformer}}. For all other baselines, we followed the suggested settings and implementations in RecBole \cite{recbole[1.0]}. To ensure a fair comparison, we conducted extensive hyperparameter tuning for each baseline method across different datasets.

\subsection{Evaluation Results} \label{subsec:eval_results}
We compared the performance of our method to baselines on four different datasets, the results are given in Table \ref{tab:all_results}. Our method achieves the best overall performance on all datasets. Notably, we observed a 9.8\% improvement in Recall@10 and a 14.4\% improvement in NDCG@10 on the footwear dataset. On sparser datasets, the gains are more significant, with our method achieving a 30.4\% improvement in Recall@10 and a 64.5\% improvement in NDCG@10 on the clothing dataset. This is because item IDs cannot capture the rich semantic relationships that are readily expressed in item texts (\textit{e.g.}, color, brand).
In comparison to Recformer, which also leveraged text information, our method has the additional advantage of incorporating general knowledge and reasoning capabilities inherent in large language models. This yielded superior performance across recommendation tasks.

\begin{table}[htbp]
\small
\centering
\caption{Ablation study of different design components.}
\label{tab:ablation_study}
\renewcommand{\arraystretch}{1.4}
\setlength{\belowrulesep}{1.4pt}
\setlength{\tabcolsep}{1.3pt}
\begin{tabular}{c|c|c|c}
\toprule
\hline
\textbf{Variants}                 & \textbf{Recall@10} & \textbf{NDCG@10} & \textbf{MRR}    \\ \hline
Ours                     & 0.2786    & 0.2037  & 0.1804 \\ \hline
w/o product attributes & 0.2293    & 0.1396  & 0.1117 \\ \hline
w/o query-product memory                & 0.2595    & 0.1786  & 0.1518 \\ \hline
w/o text embedding       & 0.1757    & 0.1569  & 0.1510  \\ \hline
w/o id embedding         & 0.2412    & 0.1480   & 0.1189 \\ \hline
w/ CLIP text embedding   & 0.2004    & 0.1245  & 0.1039 \\ \hline
w/o q.p.m. \& id emb.   & 0.2343    & 0.1424  & 0.1139 \\ \hline
w/o q.p.m. \& id. \& pro. a.   & 0.2184    & 0.1285  & 0.1006 \\ \hline
\end{tabular}
\vspace{-3mm}
\end{table}

\subsection{Ablation Study} \label{subsec:ablation_study}
We analyzed how different components in our design influence recommendation performance by introducing various model variants and testing them on the luggage and bags dataset. Specifically, we consider the following variants: (1) w/o product attributes: Product representation includes only the product title, omitting all attributes. (2) w/o query-product memory: Removes the query-product memory module. (3) w/o text embedding: Uses only ID embeddings for item retrieval. (4) w/o ID embedding: Uses only text embeddings for item retrieval. (5) w/ CLIP text embedding: Uses CLIP models for item retrieval. (6) w/o q.p.m. \& id emb.: a combination of the removals from variants (2) and (4). (7) w/o q.p.m. \& id. \& pro. a.: combining the removals from variants (1), (2), and (4).
The results in Table \ref{tab:ablation_study} show that each component improves performance. Notably, variants 3 and 4 highlight the benefits of our mixup-based retrieval method. The performance gap between variants 4 and 5 indicates that CLIP embedding models are less effective for recommendation tasks. Additionally, the slight performance drop from (4) to (6) indicates that the Query-Product Memory Module mainly influences the ID representation. Comparing (6) and (7) reveals the significance of product attributes in generating precise product titles.

\subsection{Further Investigation} \label{subsec:further_inves}

\begin{figure}[htbp]
\subfigure{
\centering
\includegraphics[width=0.465\linewidth]{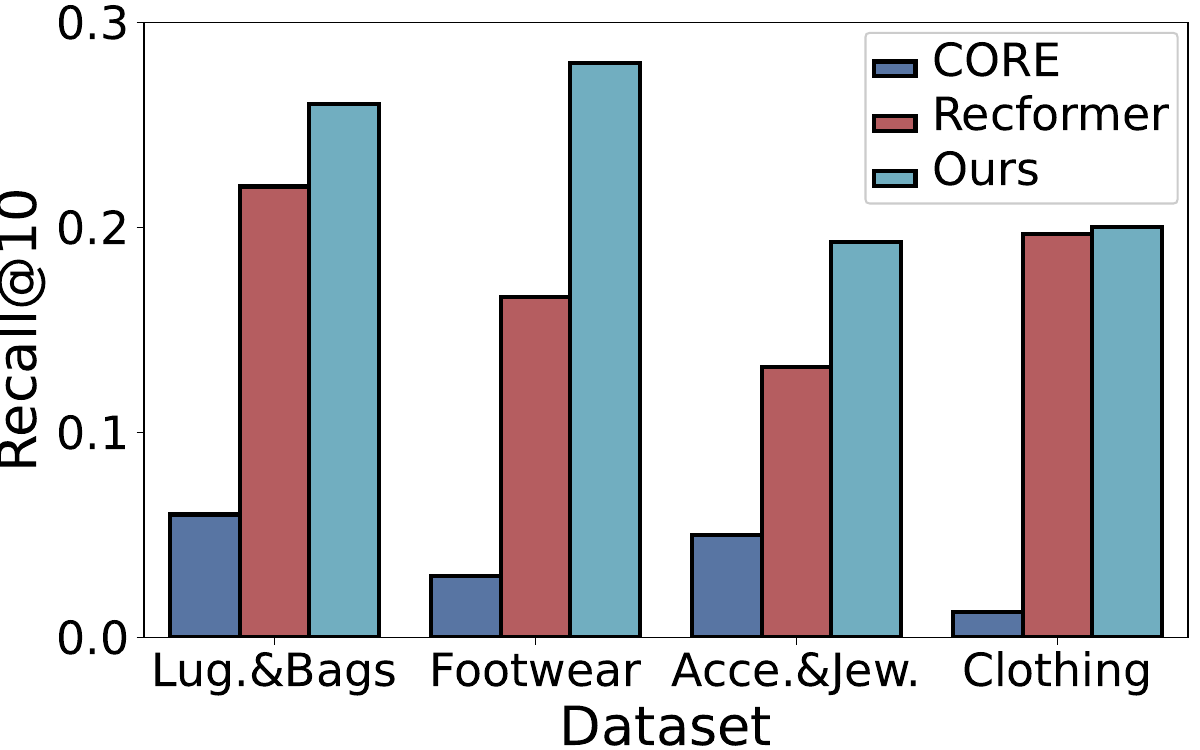}
}
\hfill
\subfigure{
\centering
\includegraphics[width=0.465\linewidth]{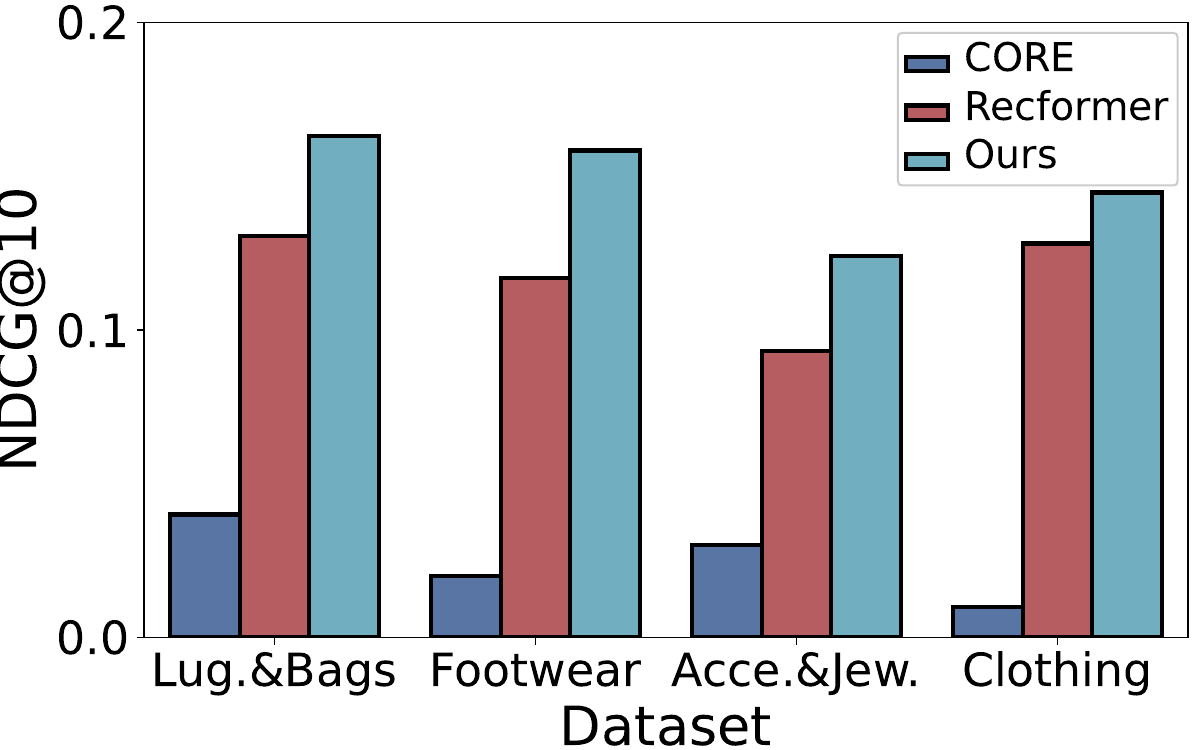}
}
\caption{Performance comparison between our method with baselines in cold-start settings.}
\vspace{-1em}
\label{fig:cold_start}
\end{figure}

\bsub{Cold-start Setting. }
The cold-start problem is a well-known issue in recommendation systems \cite{lee2019melu,pan2019warm,zhu2021learning}. To assess our model's performance in a cold-start context, we have selected items from the testing sets that have not appeared in the training sets to construct the cold-start dataset for evaluation. For ID-based methods like CORE, we incorporate a "cold" token embedding into the item embeddings to supply prior knowledge, following the approach in \cite{li2023text}. The results are presented in Figure \ref{fig:cold_start}. It is evident that text-based methods significantly outperform ID-based approaches, primarily due to the limitations of randomly initialized cold-start item embeddings. Furthermore, our method surpasses Recformer, illustrating the effective incorporation of general knowledge and reasoning capabilities provided by LLMs.

\begin{figure}[htbp]
\subfigure{
\centering
\includegraphics[width=0.465\linewidth]{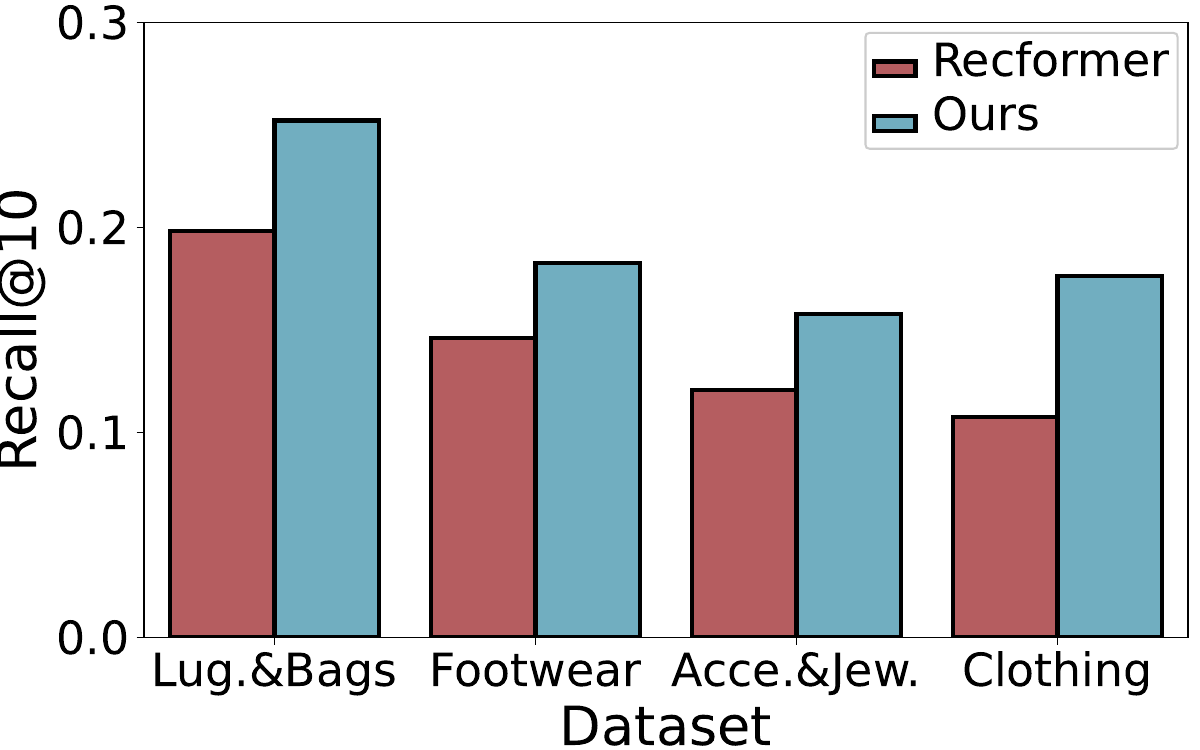}
}
\hfill
\subfigure{
\centering
\includegraphics[width=0.465\linewidth]{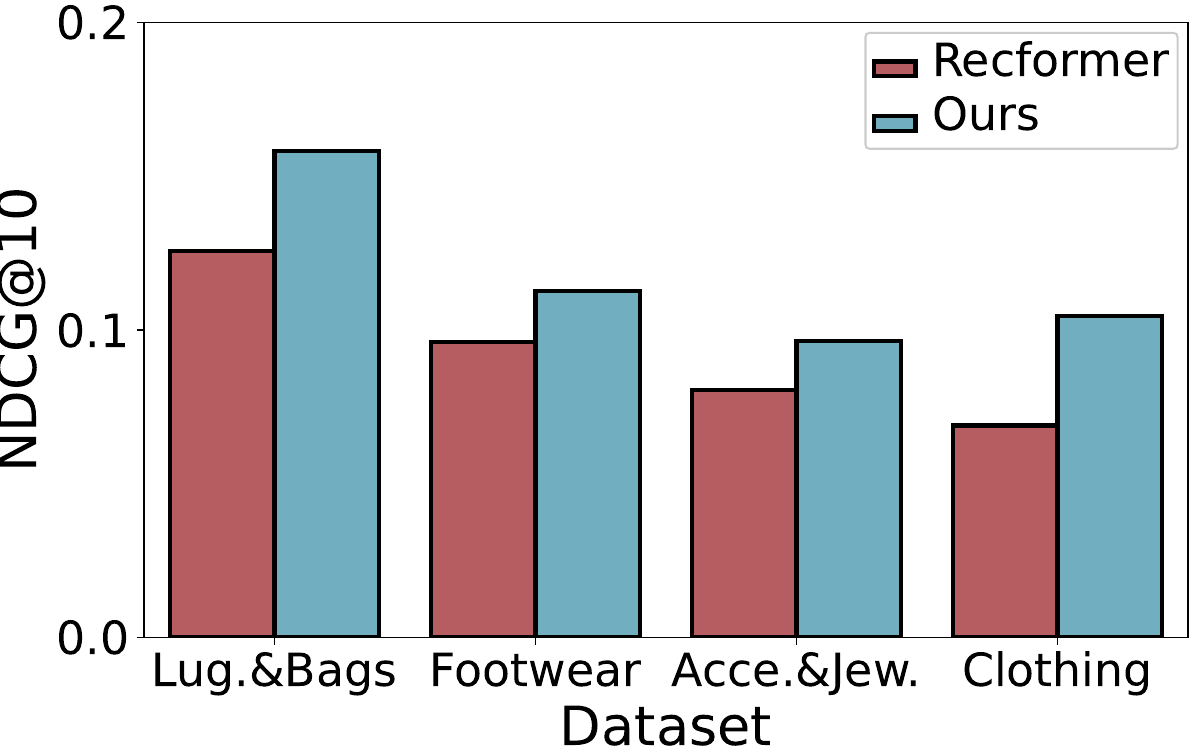}
}
\caption{Performance comparison between our method with baselines in zero-shot settings.}
\vspace{-1em}
\label{fig:zero_shot}
\end{figure}

\bsub{Zero-shot Setting. }
In this setting, the models are required to learn knowledge from pre-trained datasets and directly test on downstream datasets without further fine-tuning, thus ID-based methods are not applicable here. To ensure a fair comparison with Recformer, which undergoes pre-training on large-scale, recommendation specific datasets, we used models pre-trained on the footwear dataset to evaluate performance on three other datasets. We also employed a model trained on the luggage dataset to assess its performance on the footwear dataset. The superior performance given in Figure \ref{fig:zero_shot} demonstrates that our method can effectively capture and transfer learned knowledge to new tasks based on language understanding.

\begin{figure}[htbp]
\subfigure{
\centering
\includegraphics[width=0.465\linewidth]{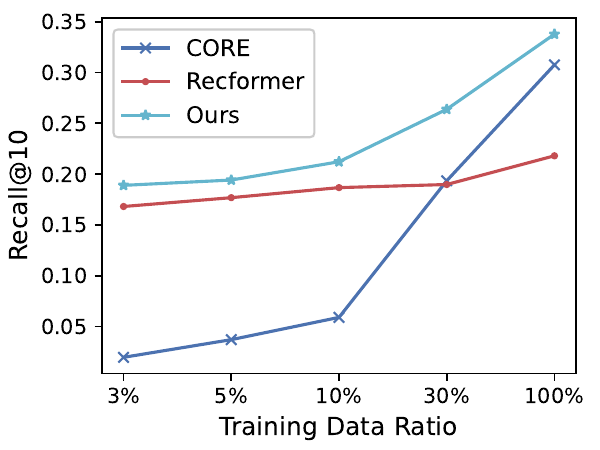}
}
\hfill
\subfigure{
\centering
\includegraphics[width=0.465\linewidth]{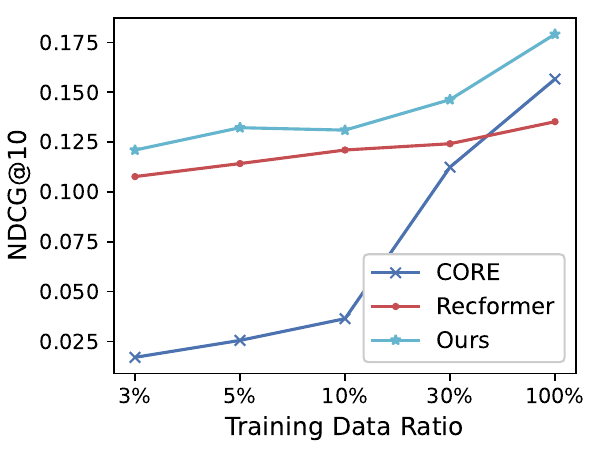}
}
\caption{Comparison between our method with baselines in low-resource settings.}
\vspace{-1em}
\label{fig:low_res}
\end{figure}

\bsub{Low Resource Setting. } 
In this setting, we trained models on datasets with different ratios of training data. The experiment results are given in Figure \ref{fig:low_res}. We can see that when the less training data is available, the text-based methods outperforms the ID-based CORE, this advantage stems from the transferable knowledge encoded in item texts. Additionally, as the amount of training data increases, our method shows a more significant performance improvement compared to Recformer, highlighting its efficiency in learning task-specific knowledge.

\section{Conclusion}
\label{sec:conclusion}

In this paper, we propose a sequential fashion recommendation system enhanced by a LLM. Our method consists of three stages: prompt creation, training and inference, and retrieval and ranking.
First, we design specialized prompts that align the model with recommendation-specific goals. Second, we conduct efficient training to optimize the model. Third, we introduce a novel mix-up-based retrieval strategy that utilizes both ID and title embeddings to finalize item recommendations.
Extensive experiments show our method significantly enhances fashion recommendation performance.

\section*{Limitations}  \label{sec:discussion}

\bsub{Training and Inference Overhead.} 
Incorporating LLMs into recommendation systems introduces additional complexities in terms of time and space. Despite these challenges, the domain of enhancing LLM efficiency is evolving swiftly, presenting strategies to alleviate these concerns. For instance, parameter-efficient fine-tuning techniques can notably reduce memory requirements and training time. In terms of inference efficiency, there is a growing body of research dedicated to developing more efficient inference frameworks. Notable contributions include LLMLingua \cite{jiang2023llmlingua}, StreamingLLM \cite{xiao2023efficient}, and PagedAttention \cite{kwon2023efficient}. These innovations demonstrate the feasibility of reducing the time and space complexities of LLMs. Furthermore, considering the substantial performance improvements the LLM could bring, the increased complexity is a worthwhile investment.

\bsub{Incorporating Visual Signals. }
Visual signals play an important role in shaping users' shopping decisions in the fashion domain. Our current approach focuses on textual data to model user interaction patterns, as incorporating images would significantly increase data collection and computational demands. However, integrating visual signals into our recommendation framework remains a promising direction. For instance, we could leverage multimodal LLMs to extract visual attributes such as color palette, lighting, textile type, shoulder style, and boot style \cite{zou2024eiven}. Incorporating these attributes into our LLM-based recommendation framework could enhance its effectiveness.

\bsub{Security and Privacy Risks.}
Like other machine learning models, LLMs are vulnerable to various security and privacy risks \cite{liu2023riatig,chang2024sok,liu2024please}. For instance, LLMs can exhibit memorization tendencies that make them susceptible to data extraction attacks, which may recover training samples and thereby compromise user privacy \cite{carlini2021extracting}. Addressing these risks with effective countermeasures is an important direction for future work.

\bibliography{reference}{}

\end{document}